# Mode I fracture toughness of asymmetric metal-composite adhesive joints

Theodoros Loutas[1*], Panayiotis Tsokanas[1], Vassilis Kostopoulos[1], Peter Nijhuis[2], Wouter M. van den Brink[2]

[1] Laboratory of Applied Mechanics and Vibrations, Department of Mechanical Engineering and Aeronautics, University of Patras, Patras University Campus, GR-26504 Rio-Patras, Greece
[2] Royal Netherlands Aerospace Centre NLR, The Netherlands
[*] Corresponding author. E-mail address: thloutas@upatras.gr (T. Loutas)

## Abstract

In this work, the mode I fracture toughness of dissimilar metal-composite adhesive joints is experimentally investigated using the double cantilever beam (DCB) test. The particular joint under study is resulted by the adhesive joining of a thin titanium sheet with a thin carbon fiber reinforced plastic (CFRP) laminate and is envisioned to be implemented in the hybrid laminar flow control system of future aircraft. Four different industrial technologies for the joining of the titanium and CFRP adherents are evaluated/compared; co-bonding with and without adhesive and secondary bonding using either thermoset or thermoplastic CFRP. The vacuum-assisted resin transfer molding (VARTM) technique is employed for the manufacturing of the panels. After manufacturing, the panels are cut into test specimens that, because they are too thin (approximately 2.4 mm thick), needed to be stiffened from both titanium and composite sides with two aluminum backing beams to ensure the non-yielding of the titanium during the subsequent DCB tests. Towards the determination of the fracture toughness of the joint from the experimental data, an analytical model recently developed by the authors, that considers the bending-extension coupling of both sub-laminates constituting the test specimen as well as the manufacturing-induced residual thermal stresses, is applied. For the four manufacturing options (MO) investigated, the load-displacement behaviors, failure patterns, and fracture toughness performances are presented and compared.

## Keywords

# 1. Introduction

Among the techniques used by the industry to join two or more parts, adhesive bonding is gaining the attention of researchers and engineers worldwide, since it offers significant advantages over other joining techniques (e.g. joining with bolts). For instance, it offers significant weight savings, uniform stress distribution in the bonding region, as well as great flexibility to design complex geometries. Adhesive bonding of similar materials (e.g. metal-metal or composite-composite joints) is well-understood and, as a consequence, several testing procedures to evaluate the fracture toughness of similar adhesive joints have been standardized.

In recent decades, dissimilar adhesive joints (e.g. metal-composite adhesive joints) are finding increasing usage in a variety of high-performance applications in several industries (e.g. aerospace, automotive, wind energy, etc.). Some typical metals used are aluminum, steel, and titanium while carbon fiber/epoxy and glass fiber/epoxy are two commonly used composites. The joining is usually achieved after curing of the adhesive in elevated temperatures, at least when the joint is intended for a high-performance application, so residual thermal stresses are inevitably generated.

The establishment of adequate methodologies for the evaluation of the fracture toughness of dissimilar adhesive joints remains a challenge due to some peculiarities that these joints have, with the most obvious to be the by definition heterogeneity in the material and/or thickness between the joint's adherents. Also, determining the fracture toughness in the presence of residual thermal stresses requires special attention and possibly calls for the standardization of some recent data reduction schemes that consider this effect.

## 1.1. Fracture toughness of dissimilar, metal-composite, adhesive joints

Intense scientific interest has recently been expressed towards the experimental investigation of the quasi-static mode I, mode II, and mixed-mode I-II fracture toughness of adhesive joints. The part of the published literature concerning similar adhesive joints (e.g. metal-metal or composite-composite joints) is intense, while the work on the fracture toughness of dissimilar adhesive joints, and more specifically of joints between metal and composite, is much more limited [1-15]. In this paragraph, we present a review of these works [1-15].

Most of the studied dissimilar adhesive joints consist of two adherents that are either isotropic or "homogeneous" (i.e. without elastic couplings), see e.g. Refs. [1-7]. In Ref. [1], new analytical expressions were derived for the computation of the SERR and the mode mixity of the mixed-mode bending test on a bi-material joint. The results were compared with finite element analyses (FEA) results of interfacial cracks in a model copper-molding joint commonly used in electronic packages. In Ref. [2], a novel idea to estimate the pure mode I interfacial fracture of adhesive joints between two dissimilar but isotropic beams was proposed. First, a simple configuration to realize nearly pure mode I fracture tests was reported. Subsequently, the concise forms of the J-integral were derived and used to characterize the interfacial fracture behavior of a dissimilar joint using the DCB test. In Ref. [3], an experimental characterization program of the fracture toughness of some composite-composite and aluminum-composite adhesive joints, bonded at various temperatures, was presented. The Euler beam theory-based analytical model used for experimental data reduction considers the temperature effect. In Ref. [4], the fracture behavior of aluminum-composite adhesive joints under DCB testing was investigated by experimental and numerical techniques, using both the virtual crack closure technique (VCCT) and the J-integral approach. The flexural rigidities of the two sub-laminates' cross-sections were required to be approximately equal so that the utilization of the modified beam theory and the compliance calibration method do not introduce significant errors. In Ref. [5], an Euler beam theory-based methodology was proposed for calculating the mode II fracture energy for adhesive joints between dissimilar materials, where the thickness of the adhesive is non-negligible compared with adherents' thicknesses. In Ref. [6], a theoretical method with parameters easily measurable from experiments was proposed to calculate the SERR of welded joints between metals and composites under DCB testing, considering the effect of manufacturing-induced thermal stresses. In Ref. [7], the problem of designing DCB tests for adhesively-bonded bi-material joints to obtain pure mode I fracture was revisited. A design criterion that requires matching of the longitudinal strain distributions of the two adherents at the bondline was presented.

Another group of works investigates the interfacial fracture of joints consisted of two "non-homogeneous" (i.e. elastically coupled) sub-laminates [8-15]. In this case, studying the various published papers [8-15], we observe the experimental data to be post-processed using different data reduction approaches, from simpler such as the Euler beam theory or William's approach to more sophisticated ones, such as the Wang and Qiao's [16] approach.

In Ref. [8], doubler plates were adhesively bonded to delamination specimens of a thin composite to prevent bending failure of the unbonded arms of the specimen. The data reduction equations for some common test configurations were re-derived for use with specimens that have bonded doublers. In Ref. [9], an experimental approach to obtaining the critical mode I and II SERR for interfacial fracture in a sandwich composite was outlined. By modifying the geometry of the sandwich beam, such that the crack plane and neutral axis coincide, the mode I and II SERR by DCB and end-notched flexure (ENF) tests, respectively, were obtained. The geometry modification required equality of the bending stiffnesses of the two arms of the beam. In Ref. [10], a semi-analytical methodology for the prediction of the fracture behavior under mixed-mode bending testing of asymmetric glass fiber reinforced plastic adhesive joints was proposed. The main advantage of that methodology is the ability to considering the fiber bridging effect as well as the arbitrariness of the adherents' stacking sequences. In Ref. [11], multi-directional fiber metal laminates were subjected to ENF tests. A methodology was then proposed to obtain the SERR and the mode mixity using an enhanced beam theory-based analytical model and was validated by the standardized compliance calibration method. In Ref. [12], plate theory analyses were employed to obtain the SERR and the mode mixity of DCB tests on carbon fiber aluminum laminates and were compared with the compliance calibration method's predictions. The SERR acquired by both methods were identical in the initial crack length but by increasing it, the fracture energies estimated by the plate theory surpassed the compliance calibration method counterpart. In Ref. [13], the mode I fracture toughnesses of the metal/composite interface region of some fiber metal laminates were determined and a finite element model was developed to account for the influence of metal plasticity on the measured fracture toughnesses. In Ref. [14], modified DCB specimens were tested to investigate the mode I fracture properties of an asymmetric metal-composite adhesive joint. A modified Kanninen theory was then used to consider the specific specimen design. In Ref. [15], the interlaminar fracture toughnesses of some glass laminate aluminum reinforced epoxy (so-called GLARE) laminates were investigated by simple but approximate beam theory- and fracture mechanics-based analytical solutions.

## 1.2. Current work

In the present work, the quasi-static mode I interfacial fracture toughness of adhesively bonded joints between titanium and CFRP is experimentally investigated using the DCB test configuration. Both titanium and CFRP adherents of the present joint are very thin, thinner than 1.5 mm, as the aircraft application for which it is intended requires [17]. In our latest paper [18] we propose an engineering approach for the design of fracture toughness tests (i.e. DCB and ENF tests) for the present adhesive joint.

Here, four different industrial technologies (hereinafter refer to as MO) are considered for the manufacturing of the titanium-CFRP joint; namely co-bonding with and without adhesive, secondary bonding using thermoset composite, and secondary bonding using thermoplastic composite. Thus, each MO uses different composite materials, adhesive agents, and/or bonding technologies. All MO require the joint to be subjected to high temperature during its manufacturing, which generates residual thermal stresses. During the manufacturing stage, after production of the panels and extraction of test specimens from them, since both adherents are too thin, the specimens are backed from both titanium and composite sides with two aluminum stiffening beams (Figure 1a) to prevent large/plastic deformation of the crack arms during the subsequent tests [18]. With an aim to investigate the interfacial fracture behavior of the backed and with residual thermal stresses joint, quasi-static DCB tests are performed at room temperature conditions and as close to the requirements of the ASTM standard as possible. Post-mortem fractographic analyses are undertaken to gain further insight into the involved fracture mechanisms. The data reduction scheme we recently proposed in Ref. [19] is used for experimental data reduction; i.e. calculation of the SERR and mode partitioning. As Ref. [19] details, that data reduction scheme can consider both the bending-extension coupling induced by the presence of the aluminum beams and the manufacturing-induced residual thermal stresses. The four MO studied are compared in terms of their load-displacement responses, failure patterns, and fracture toughness performances.

## 2. Materials and methods

### 2.1. The manufacturing options (MO) under investigation

The following four MO, also schematically presented in Figure 1b, are evaluated in the present work:

- MO 1: Vacuum infusion/resin transfer molding (RTM) of the thermoset CFRP plate followed by secondary bonding of it to the titanium sheet using film adhesive (FM 94K).
- MO 2: Co-bonding using film adhesive (FM 300M) on the interface between titanium and thermoset CFRP and vacuum infusion/RTM.
- MO 3: Co-bonding without the use of adhesive on the interface between titanium and thermoset CFRP and vacuum infusion/RTM. The bonding is achieved by the excess RTM6 resin of the CFRP.
- MO 4: Secondary bonding of the thermoplastic CFRP to the titanium sheet using film adhesive (FM 94K).

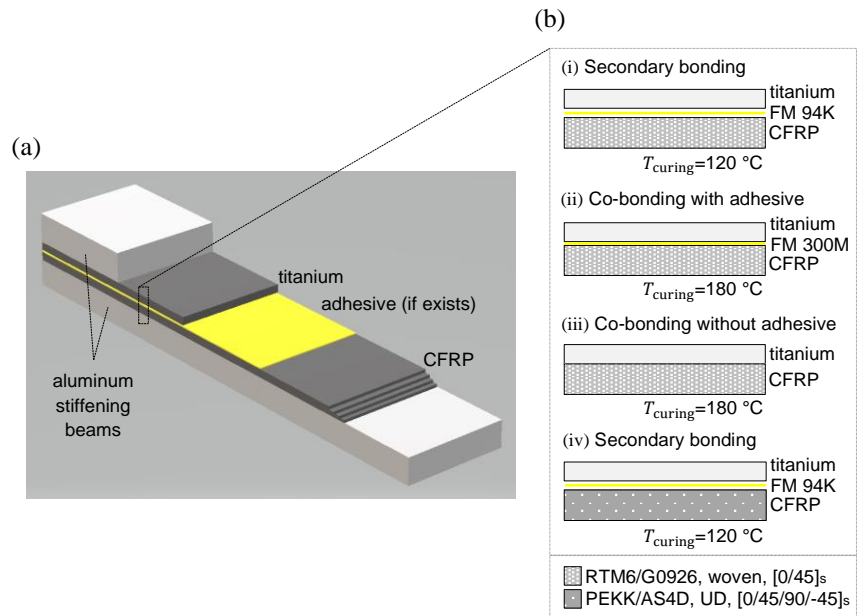

Figure 1: The titanium-CFRP adhesive joint under investigation. (a) The sequence of the individual layers constituting the complete adhesive joint, after the addition of the aluminum stiffening beams. (b) Schematic representation of the four manufacturing options (MO) under study; (i) MO 1, (ii) MO 2, (iii) MO 3, and (iv) MO 4.

### 2.2. Materials

The materials used are the following:

- Titanium Grade 2 (CP40), with the rolling direction parallel to the length direction of the resulting test specimens.

- For the thermoset CFRP: HexFlow RTM6 epoxy resin (from Hexcel) and 5-harness weave fabric Hexforce G0926 (from Hexcel) with a 6K HS carbon fiber and an areal weight of 370 gsm.
- For the thermoplastic CFRP: Cetex TC1320 PEKK (from TenCate) with an AS4D fiber and an areal weight of 145 gsm.
- Aluminum 2024 T3.
- Adhesives.
    - FM 94K 0.06 psf adhesive film (from Solvay) with knit carrier, areal weight equal to 293 gsm, and nominal thickness equal to 0.25 mm.
    - FM 300M 0.03 psf adhesive film (from Solvay) with mat carrier, areal weight equal to 150 gsm, and nominal thickness equal to 0.13 mm.

For the needs of the present work, the following materials are also used: (a) 3M Scotch-Weld 9323 B/A adhesive with a nominal thickness equal to 0.20 mm, and (b) Upilex-25S foil with a thickness equal to 0.025 mm.

As understood, we use two different aerospace grade epoxy-based adhesives for the joining of titanium and CFRP, namely the FM 94K and FM 300M adhesives, considering the different curing temperature that each MO requires as well as the compatibility between the candidate adhesive and the composite. Specifically, for secondary bonding, an adhesive with low curing temperature is preferred to minimize the thermal stresses in the materials. Also, we preferred to use a knitted carrier, to create an even adhesive thickness of around 0.2 to 0.3 mm. For co-bonding, an adhesive with a similar curing temperature to the CFRP, in our case 180 °C, is preferable. Also, a thinner adhesive with a mat carrier can be used.

In Table 1, the material properties and thicknesses of the utilized materials are summarized.

Table 1: Engineering constants, coefficient of thermal expansion (CTE), and nominal thicknesses of the constituent materials of the titanium-CFRP adhesive joint.

| Material | $E_1$ (GPa) | $E_2$ (GPa) | $G_{12}$ (GPa) | $v_{12}$ (-) | CTE $\alpha_1$ ($\cdot 10^{-6}$/°C) | Thickness (mm) |
|---|---|---|---|---|---|---|
| Titanium | 105.0 | - | 45.0 | 0.340 | 8.6 | 0.800 |
| CFRP, woven[1] | 66.0 | 66.0 | 4.5 | 0.035 | 2.9 | 0.363 |
| CFRP, UD[1] | 139.0 | 10.5 | 5.2 | 0.076 | 1.0 | 0.140 |
| Aluminum | 73.1 | - | 28.0 | 0.330 | - | 5.000 |

[1] The given properties and thicknesses refer to the level of layer
CTE: coefficient of thermal expansion

## 2.3. Manufacturing processes

Since the under-study adhesive joint is envisioned to be applied on an aircraft leading edge [17], the processes we use here to manufacture the joint are also applicable by the aerospace industry[1]. Some representative "snapshots" from the manufacturing processes are shown in Figure 2.

### 2.3.1. Manufacturing following the manufacturing option (MO) 1

The MO 1 panel was produced in two steps; first, the vacuum injection of the composite laminate and, after curing and quality assessment of it, the bonding of the surface treated titanium sheet. The vacuum injection was done on a flat oil/water heated mold using, as schematically presented in Figure 2a, a double vacuum bag, i.e. one vacuum bag covering the composite layers and flow media and a second one covering the first vacuum bag and the caul plate. Since the same process was also used for the production of the panels from the MO 2 and 3 (see Paragraphs 2.3.2 and 2.3.3, respectively), the flow media were chosen to be positioned on the mold side, so for the needs of these MO, the caul plate can be replaced by the titanium sheet on the outside.

---

[1] The envisioned aircraft leading edge [17] would be manufactured with the composite towards a male mold and the titanium on the outside towards the vacuum bag.

The resin was injected at a temperature of 80 °C and a mold temperature of 90 °C. After injection of the correct amount of resin to achieve a 57% fiber volume fraction, the injection stopped and the curing cycle started. The composite laminate was cured at a temperature of 180 °C for 1.5 h, applying a ramp-up rate of 1.5 °C/min and a cool down rate of 0.5 °C/min. After the curing process, the composite laminate was inspected using C-scan.

The titanium sheets were surface treated before bonding using methods described in a previous paper [20]. The treatment consisted of PFQD solvent cleaning, grit blasting at 3 bar pressure, PFQD cleaning, UV/Ozone treatment, AC-130 Sol-gel application, and BR6747-1 primer application.

The titanium sheets were bonded to the composite laminate in an autoclave. After the pre-treatment process and just before the bonding, the titanium sheets were cleaned with alcohol. FM 94K adhesive was positioned on the titanium surface, together with a Upilex foil used to create an artificial crack. No adhesive was used at the locations of the Upilex foil, i.e. in the interfaces between Upilex and titanium or composite. The Upilex was Frekoted at its both sides.

Before positioning the composite panels, they were treated with Ozone. The following steps were done just before closing the autoclave bagging: removal of peel ply, cleaning with acetone, grit blasting with Corundum (aluminum oxide) with 2 bar pressure, cleaning with acetone, and treatment with UV/Ozone light for 7 min.

After the treatment, a standard autoclave bag was made to cure the adhesive. No more than 2 h passed between Ozone treatment and the start of the autoclave cycle. The adhesive was cured for 1 h at 120 °C using ramp up and cool down rates of 2 °C/min, as well as applying a pressure of 1.8 bar and full vacuum.

Since the available equipment for the surface treatment of the titanium sheets was limited in size, the joints from all MO were made of six titanium sheets in one composite laminate, as can be seen in Figures 2b and 2c.

### 2.3.2. Manufacturing following the manufacturing option (MO) 2

The manufacturing of the MO 2 panels was identical to that of MO 1 panels, but instead of using a peel ply and a caul plate on top of the composite layers, we used an adhesive agent, Upilex, and a titanium sheet. Thus, no separate bonding step was needed since both resin and adhesive were cured during the curing cycle of the vacuum infusion process. A snapshot during manufacturing is shown in Figure 2b in which some titanium sheets under the vacuum bag can be seen. After removing the vacuum bag, the co-bonded titanium-CFRP panel was found curved, as Figure 2c shows. The panel was cut into smaller parts to enable its C-scan inspection in our facilities.

It is noted that the panels and subsequent test specimens from all MO appeared to be curved, which is evident as they all consist of two adherents with different coefficients of thermal expansion and were manufactured (co-cured or co-bonded, depending on the MO) in high temperature (120 °C or 180 °C, depending on the MO).

### 2.3.3. Manufacturing following the manufacturing option (MO) 3

The manufacturing of the MO 3 panels was identical to that of the MO 2 ones, except that no adhesive film was used here. Unfortunately, during the injection, the amount of resin injected was slightly too low because the heating of the injection hose accidentally broke down and, thus, the injection hose had to be replaced during the injection. By this, the resin weight measurement was disturbed and, as a result, some "dry spots" were created, as demonstrated by the respective C-scan image (see Figure 2d).

Especially for the MO 3 panels, the C-scan inspection process we followed is schematically presented in Figure 2e. The panels were first scanned from the composite side. Next, to see if the areas with high attenuation were caused by excessive porosity or possible disbonding in the titanium/CFRP interface, the specimens cut from the panels (see Paragraph 2.3.5 for more information on the specimens cutting) were inspected again, this time from the titanium side. Although no disbondings were found, the specimens extracted from the regions of the panels with high attenuation were still not used for the tests, since the dry spots could still influence the results of the DCB tests.

### 2.3.4. Manufacturing following the manufacturing option (MO) 4

For the needs of the MO 4, the CFRP laminates were produced by fiber placement followed by consolidation in an autoclave.

The preform was prepared using an in-house fiber placement machine with laser heating. After fiber placement, the preform was consolidated in a normal high-temperature vacuum bag with a caul plate. The composite was heated up to 385 °C for 45 min with a ramp-up rate of 3 °C/min and a cool down rate of 5 °C/min. The pressure was increased from 2 to 7 bar after reaching a temperature of 250 °C and full vacuum was applied to the system. After consolidation, the composite was inspected with C-scan and, then, the titanium sheets were bonded to the composite as in the case of the MO 1 panels, including all the pre-treatment processes mentioned in Paragraph 2.3.1.

It is noted that for the MO 4, we chose the secondary bonding technique because otherwise, i.e. if we had performed co-bonding, we would have introduced extremely high residual thermal stresses in our joint.

We note that for the MO 1 and 4, we performed C-scan inspection both before and after the secondary bonding of the titanium sheets.

### 2.3.5. Preparation of the test specimens

After completion of the manufacturing and quality assessment of the panels, we cut them to the desired dimensions using a waterjet cutter to create test specimens. The cutting was performed starting from the titanium surface and specifically from the areas without Upilex to prevent possible delaminations in the CFRP. Unfortunately, since the panels were curved, the cutting was sometimes disturbed and, as a result, some of the specimens did not have a constant width. Those specimens were not used in the experiments.

Aluminum backing beams were adhesively bonded on the top and bottom surfaces of the specimens, as shown in Figure 1a. The bonding was undertaken after cleaning with alcohol, sanding with sanding paper 120, and cleaning again the surfaces to be bonded. As aforementioned, before backing, the specimens produced following all MO were curved (see Figure 2c-ii) and, thus, to achieve the bonding of the aluminum backing beams, they were pressed flat. M3 screws were installed in the M3 holes in the aluminum backing beams (the holes are schematically shown in Figure 3) to prevent the adhesive from flowing into the holes. Glass pearls with a diameter between 0.2 and 0.3 mm were added to the adhesive and used as spacers between specimen's surface and aluminum, to get the desired adhesive thickness. The alignment of the aluminum backing beams with the titanium-CFRP specimens was performed manually using a flat table and some small rectangular blocks, without using any special equipment.

After bonding, the adhesive was cured for 24 up to 48 h at room temperature, so the aluminum bonding process did not introduce additional thermal stresses to the joint.

The total thickness of the final, backed, joint is approximately 12 mm, with small differences between the four MO.

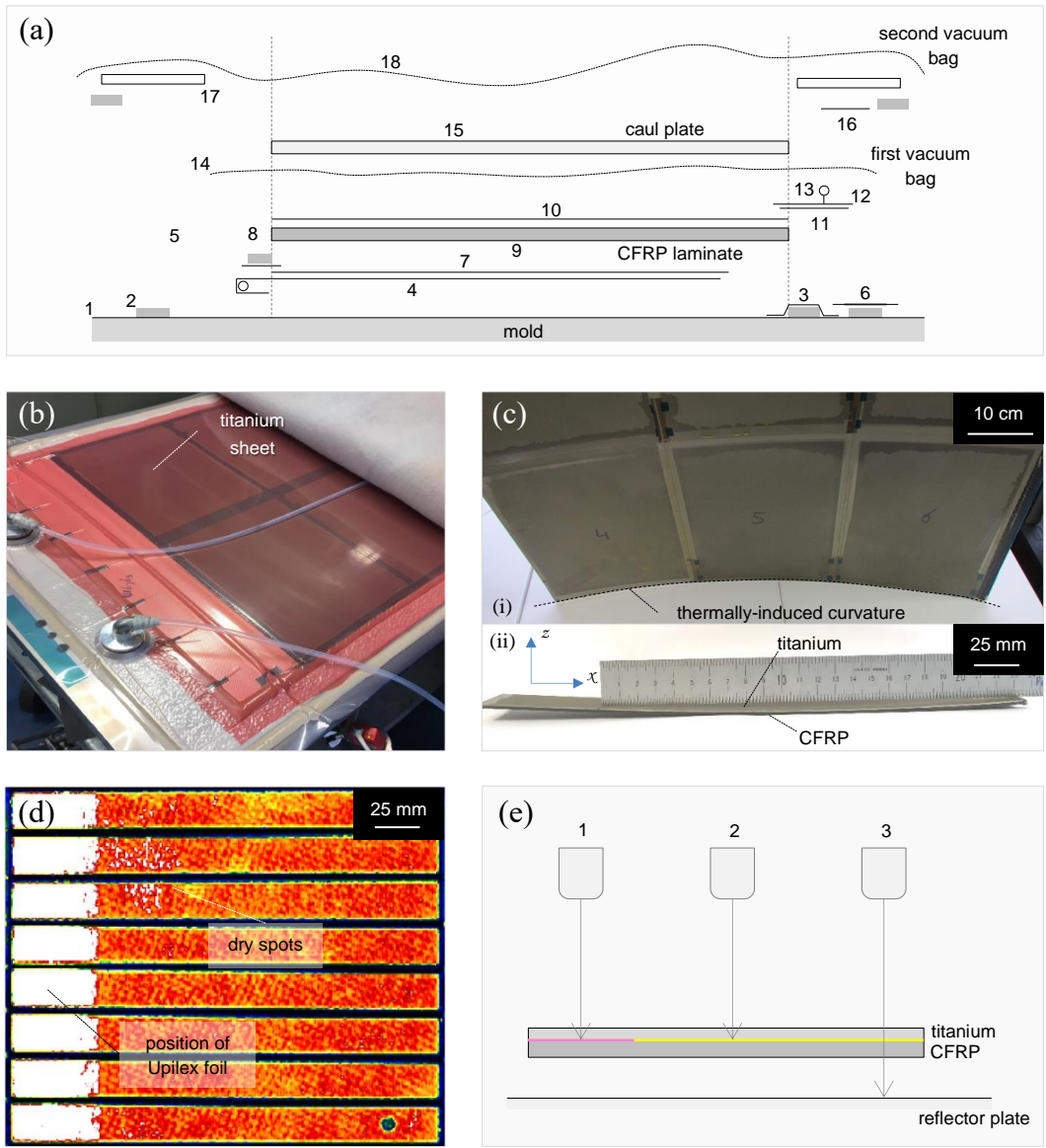

Figure 2: (a) Vacuum bag schematic overview. [(1) Heated mold. (2) Tacky tape. (3) Tacky tape covered with Flashbreaker tape. (4) Flow mesh. (5) Injection tube. (6) Glass fiber breather strings with Flashbreaker cover. (7) Porous PTFE. (8) Tacky tape on top of Flashbreaker tape. (9) Composite layers. (10) Peel ply. (11) Porous PTFE. (12) Glass fabric. (13) Exit tube. (14) Vacuum foil. (15) Caul plate. (16) Flow mesh under vacuum point. (17) Airweave. (18) Vacuum foil.] (b) A photograph during the production of the panels via the vacuum-assisted resin transfer molding (VARTM) process. (c) Produced titanium-CFRP adhesive joint. All panels (i) and test specimens (ii) cut from them appeared to be curved, due to manufacturing-induced residual thermal stresses. (d) Using C-scan inspection, some dry spots were found on the panels produced following the manufacturing option (MO) 3, due to the relatively low amount of resin injected, caused by a mistake during the production process. (e) Schematic representation of the quality assessment of the MO 3 panels via the C-scan technique; (1) back-wall C-scan, (2) C-scan of the titanium/CFRP interface, and (3) attenuation C-scan (reflector plate), titanium side.

## 2.4. Mechanical experiments

The quasi-static mode I interfacial fracture toughness of the under-study titanium-CFRP adhesive joint was experimentally measured using the DCB configuration. The experiments were performed at a 25 kN Instron 8872 universal testing machine at room temperature conditions (25 °C and 50-60% RH), following the general guidelines of the ASTM D 5528-01 test standard.

The experimental setup is schematically presented in Figure 3. The basic dimensions of the DCB test configuration, namely the width b, total length L, and initial crack length $a_0$, are shown in this figure while their values are given in Table 2. The position of the Upilex foil between titanium and CFRP is also shown in the same figure. The utilized piano hinges, screwed onto the aluminum beams as Figure 3 shows, were stiffer than those "typically" used in DCB experiments, because of the high expected loading values during the tests.

As shown in Table 2, the $a_0$ is not the same for all MO. Our original intention was to create a natural crack before starting the test because, as known, a natural crack has a sharp crack tip, leading to more accurate fracture toughness values and is generally recommended in literature instead of using starter films that introduce blunt crack tips. Nevertheless, in MO 1 and 2 cases, in every attempt we made to introduce a natural crack, applying either opening (mode I) or wedge loading to the specimen, we always got delamination inside the composite. Thus, we decided to start the propagation from the Upilex foil insert right away ($a_0$=28.0 mm). For the MO 3 specimens, each experiment was performed in two steps. First, opening loading was applied to propagate the crack until the $a_0$ to become approximately 70 mm and, then, the specimen was completely unloaded. In the second step, the specimen was re-loaded until the crack length was progressed to an additional length of approximately 40 mm. For the MO 4 specimens, wedge loading was first applied to the specimens to create a natural pre-crack. The new initial crack length, i.e. from the loading axis to the crack tip of the natural crack, was measured and is the one used in the post-processing of the experimental data (see Paragraph 2.5). By the wedge loading, being an abrupt loading type in nature, it was not easy to control the propagated crack length, so the four MO 4 specimens have different $a_0$ values, as shown in Table 2.

During the tests, the specimens were loaded in tension at a crosshead velocity of 10 mm/min. The applied load P and load-point displacement δ were continuously recorded during the test. The load-displacement curves were registered for the posterior evaluation of the fracture response of the joint.

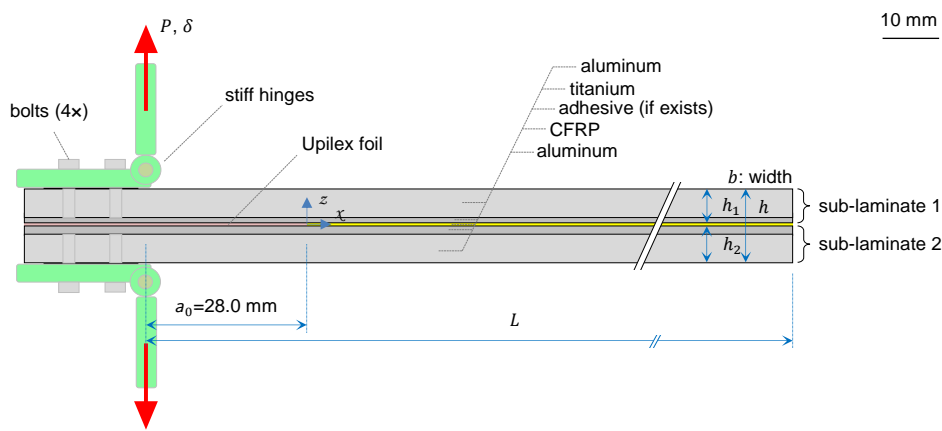

Figure 3: Experimental setup (in scale) for the double cantilever beam (DCB) experiments.

Table 2: Dimensions of the double cantilever beam (DCB) test configuration.

| MO | b (mm) | L (mm) | $a_0$ (mm) |
|---|---|---|---|
| 1 | 25 | 200 | 28.0 |
| 2 | 25 | 200 | 28.0 |
| 3 | 25 | 267 | 70.0 |
| 4 | 25 | 200 | 34.5, 54.5, 28.0, and 67.6[1] |

[1] These values correspond to the Exp. #1, #2, #3, and #4, respectively (see Figure 5d)
MO: manufacturing option

## 2.5. Experimental data reduction

The DCB specimens, schematically presented in Figure 3, consist of two sub-laminates that exhibit bending-extension coupling as well as are stressed by residual thermal stresses (Figure 2c) due to the manufacturing at high temperature. The analytical model reported in Ref. [19], as opposed to common analytical models used for data reduction from fracture toughness tests, considers both the bending-extension coupling and residual thermal stresses effects. Thus, it is utilized in the present work to estimate the total SERR of the titanium-CFRP adhesive joint, as well as the "parasitic" mode mixity due to the residual thermal stresses.

As shown in Figure 4, the two-dimensional analytical model considers an elastic, laminated, and cantilever beam with an arbitrary stacking sequence, having an asymmetric delamination crack at its left end, being loaded at the same end with concentrated vertical loads, and containing residual hygrothermal stresses. The delamination splits the laminated beam into two beams, the upper one (named "sub-laminate 1") and the lower one (named "sub-laminate 2"), both of which have arbitrary stacking sequences and are modeled as Timoshenko beams.

The conditions required to satisfy the continuity of displacements at the crack tip and along the bonded part of the beam were provided by the so-called in literature "semi-rigid interface joint model" [16]. The governing equation of the mathematical problem is a second-order non-homogeneous ordinary differential equation, whose solutions give the internal loadings $\mathcal{N}_i$, $\mathcal{Q}_i$, and $\mathcal{M}_i$ along the length x of the undelaminated portion of the beam (i.e. from section C to section D in Figure 4). The loads applied to the crack-tip element are already determined by a global beam analysis; the forces and moment at $x = 0^-$ are resulted from simple transfer of the loads P while at $x = 0^+$ are defined in Ref. [19]. The crack-tip forces $\mathcal{N}_c$ and $\mathcal{Q}_c$ are obtained by imposing static equilibrium at the crack tip (Figure 4). Following Irwin's approach, the mode I, mode II, and total SERR can be expressed in terms of the crack-tip forces ($\mathcal{N}_c$ and $\mathcal{Q}_c$) and two flexibility coefficients, i.e. two parameters that are functions of the geometric and elastic properties of the crack-tip element.

For the DCB test configuration, the following expressions for the mode I, mode II, and total SERR are used [19]:

$$G_I = \frac{1}{2}(c_1 + c_2)\left\{P(1+\lambda a_0) + \frac{2\lambda\xi\left[\alpha_{N2}-\alpha_{N1}+\frac{\eta}{\xi}(\alpha_{M2}-\alpha_{M1})+\frac{h_1+h_2}{2}\alpha_{M2}\right]}{2(d_1+d_2)\eta+[2b_1+2b_2+(h_1+h_2)d_2]\xi}\right\}^2,$$

$$G_{II} = \frac{1}{2}\left(a_1 + a_2 - h_1 b_1 + h_2 b_2 + \frac{h_1^2}{4}d_1 + \frac{h_2^2}{4}d_2\right)\left\{\frac{2}{h_1\xi+2\eta}\left[-\xi P a_0 - \alpha_{N1} + \alpha_{N2} + \frac{h_1}{2}\alpha_{M1} + \frac{h_2}{2}\alpha_{M2}\right]\right\}^2, \text{ and}$$

$$G = G_I + G_{II}.$$

(1)

In Eqs. (1), $G_I$, $G_{II}$, and G are the mode I, mode II, and total SERR, respectively. P is the applied load in the DCB test, $a_0$ is the initial crack length of the beam, and $h_i$, i=1, 2, is the thickness of the sub-laminate i (Figure 3). $a_i$, $b_i$, $c_i$, and $d_i$, i=1, 2, are the extensional compliance, bending-extension coupling compliance, shear compliance, and bending compliance of the sub-laminate i. $\alpha_{Ni}$ and $\alpha_{Mi}$, i=1, 2, are the axial strain and curvature of the sub-laminate i due to residual hygrothermal stresses. $\lambda$, $\xi$, and $\eta$ are auxiliary parameters, functions of the $a_i$, $b_i$, $c_i$, $d_i$, and $h_i$, i=1, 2 [19]. It is highlighted that the Eqs. (1) are valid only in the case of linear elastic fracture mechanics (small fracture process zone).

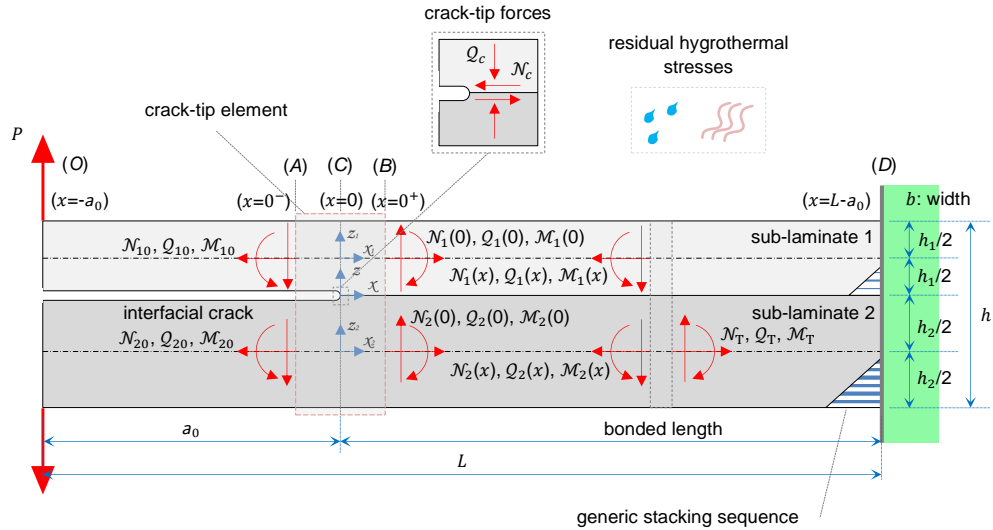

Figure 4: Schematic representation of the beam model [19] used for the determination of the fracture toughness of the titanium-CFRP adhesive joint after the double cantilever beam (DCB) experiments. An elastic, laminated, and cantilever beam, consisting of two sub-laminates with arbitrary stacking sequences as well as an asymmetric through-the-width interfacial disbonding, is loaded with two opposite loads of intensity P at its left end as well as it contains residual hygrothermal stresses. b, L, and $a_0$ are the width, total length, and initial crack length of the beam, respectively. $h_i$, i=1, 2, is the thickness of the sub-laminate i. $\mathcal{N}_c$ and $\mathcal{Q}_c$ are the crack-tip forces. The internal forces and moment, $\mathcal{N}_i$, $\mathcal{Q}_i$, and $\mathcal{M}_i$, i=1, 2, developed at various cross-sections of the beam, are shown.

## 3. Results and discussion

### 3.1. Load-displacement curves

In Figure 5, the load-displacement curves from the DCB experiments are summarized. Every curve corresponds to one successful experiment while the four different diagrams serve to compare the fracture behaviors of the four MO studied.

As shown in Figure 5a, the specimens from the MO 1 initially perform a linear load-displacement behavior, followed by a visual deviation from linearity that takes place approximately 200 N before reaching the maximum test load. After that load, a sudden load drop occurs in all four specimens tested and is associated with the abrupt propagation of the crack (see Paragraph 3.2).

Regarding the MO 2 specimens (Figure 5b), an almost linear load-displacement behavior is observed up to the maximum load, followed by the first load drop. After that drop, the load starts to increase again, this time in a strongly non-linear fashion, until it increases by approximately 200 N, so a sudden load drop occurs.

As regards MO 3 (Figure 5c), the initial portion of the curves can be characterized as approximately linear and, after reaching a maximum load of about 400 N, the load starts to decrease as the delamination propagates. It is noted that the slight deviation in the slope of the linear portion of the curves for the four specimens tested, something not observed in the previous two sets of curves, originates from the slightly different values of the initial crack length (see Paragraph 2.4). At the crack propagation phase, the curves show a mild "saw-toothed" pattern, indicating a brittle behavior of the interface.

As already mentioned, each MO 4 specimen has a different initial crack length and, as a result, each curve has a linear portion of a different slope as well as a different maximum load. As shown in Figure 5d, with the increase of the specimen's initial crack length, both the slope of the linear portion of the curve and the maximum load decrease. All MO 4 specimens exhibit an unstable crack growth characterized by sudden load drops.

Obviously, the load-displacement curves of the four MO (Figure 5) are not directly comparable to each other, since they do not correspond to the same initial crack length. For this reason, the initial crack length to which each diagram/curve corresponds is given in Figure 5, so that the reader can perform valid comparisons.

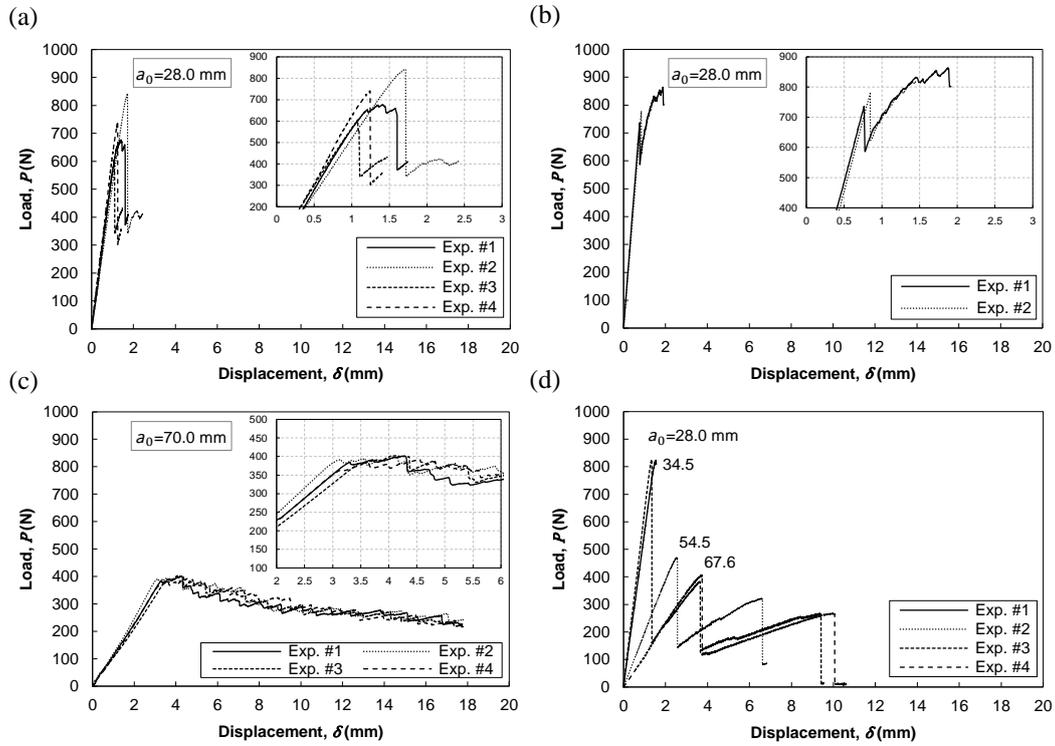

Figure 5: Load (P) versus crack opening displacement (δ) curves from the double cantilever beam (DCB) experiments, for the four manufacturing options (MO) under study; (a) MO 1, (b) MO 2, (c) MO 3, and (d) MO 4. $a_0$ is the initial crack length.

### 3.2. Fracture behaviors during testing

For the specimens from the MO 1 and 2, crack propagation monitoring using a high-resolution camera showed the development of a secondary, interlaminar in the composite, crack, started just after the initiation of the interfacial disbonding on the adhesive layer (Figure 6a). Then, the length of the secondary crack continued to increase along with that of the primary crack, until the end of the test. Based on this behavior, it seems that the pre-treatment performed [20] was good enough so that the strength of the titanium/CFRP interface was high and, subsequently, the CFRP itself became the weakest "link" in the joint. Secondary cracking was also observed in the MO 3 specimens. In this case, too, the development of the secondary crack started just after the initiation of the primary crack.

In Figure 6b, photographs of the fracture surfaces of one of the MO 4 specimens are shown and correlated with the respective load versus displacement data presented in the previous paragraph. As seen in this figure, the regions of the fracture surfaces corresponding to the "crack arrest" phases of the test appear to be rough, most likely due to the plastic deformation of the adhesive layer in those regions. On the contrary, the regions corresponding to the fast propagation phases of the test are smooth.

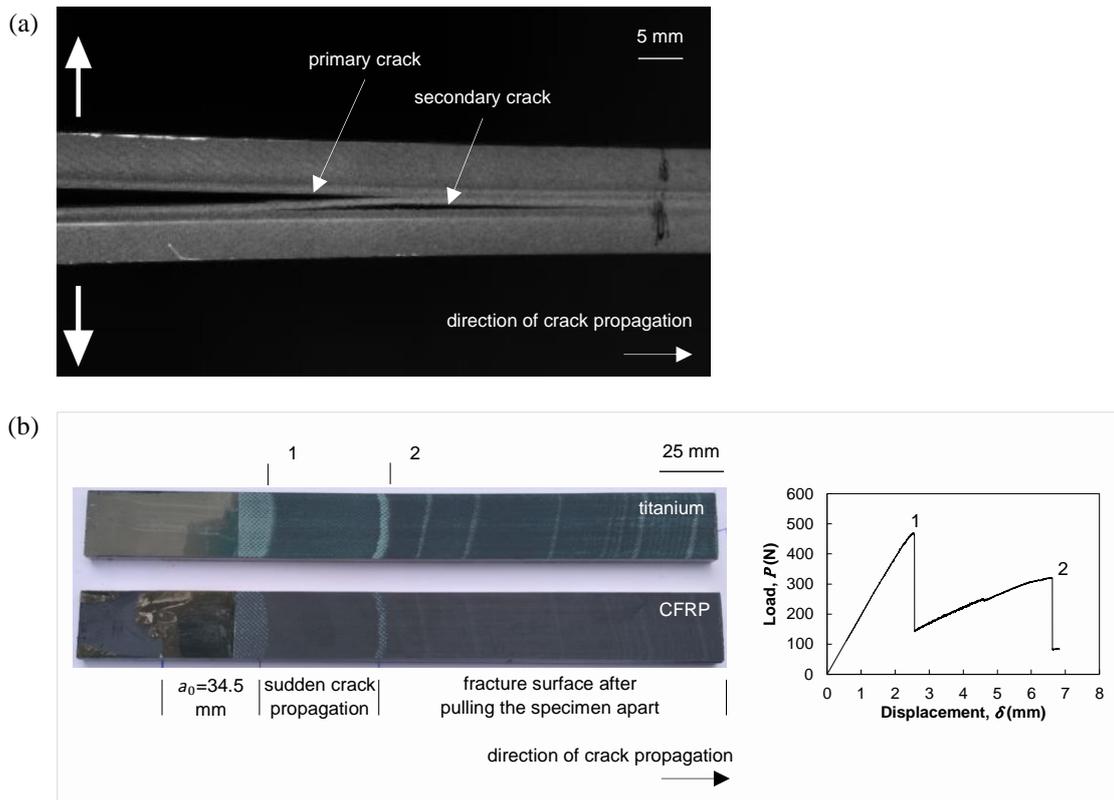

Figure 6: (a) A representative snapshot capturing the development of a secondary crack during the experimental testing of the specimens from the manufacturing options (MO) 1 and 2. (b) Fracture surfaces of a typical specimen from the MO 4.

### 3.3. Fracture toughness performances

This section concludes with the calculation of the total SERR and mode mixity of the under-study titanium-CFRP joint, utilizing the experimental data already presented, to compare the fracture toughness performances of the four MO under consideration.

Table 3 summarizes the results. As shown in this table, in terms of the total SERR, the best performing MO is the MO 4 while the worst ones are the MO 1 and 2. The "parasitic" mode mixity induced remains for all MO quite low at the loads that crack initiation occurs.

Table 3: Total strain energy release rate (SERR) (G) and mode mixity ($G_{II}/G$) values for the four manufacturing options (MO) under study.

| MO | Load, $P_{ini}$ (N) | Initial crack length, $a_0$ (mm) | SERR, G (N/m) | Mode mixity, $G_{II}/G$ (%) |
|---|---|---|---|---|
| 1 | 711 (±77) | 28.0 | 477.0 (±136) | 3.7 |
| 2 | 739 (±23) | 28.0 | 467.0 (±32) | 6.1 |
| 3 | 380 (±10) | 70.0 | 683.1 (±41) | 5.3 |
| 4 | - | - | 874.4 (±134) | 4.1 |

MO: manufacturing option. SERR: strain energy release rate

# 4. Summary and conclusions

The present work aimed to investigate the mode I fracture behavior of an adhesive joint between two aerospace grade titanium and CFRP adherents. The industrial end-user of the joint (i.e. Aernnova) required that the adherents' thicknesses are very small, smaller than 1.5 mm. Consequently, aluminum backing beams were added to avoid large deformation of the adherents (or even plastic deformation of titanium) during the tests, a prerequisite for extracting correct fracture toughness properties. In our recent papers [17, 18] we present the design of fracture toughness tests on the present adhesive joint, as well as its envisioned application in future aircraft.

Here, four different MO were explored using cost-effective industrial manufacturing processes. These processes were co-bonding with and without adhesive and secondary bonding using either a thermoset or a thermoplastic composite. Some of the specifications of the standardized DCB test were modified to accommodate the large thickness of the complete joint. An analytical model recently developed by a sub-set of the present authors, that considers the bending-extension coupling of both sub-laminates of the joint as well as the manufacturing-induced residual thermal stresses effect, was used for post-processing of the experimental data.

The main conclusions drawn from the present study may be summarized as follows:

- For the MO 1, 2, and 3, the adhesive (for the MO 1 and 2) or non-adhesive (for the MO 3) interface between titanium and CFRP is of enhanced strength with the pre-treatment techniques applied. Thus, the composite itself appears to be the "weak link" of the joint and, as a result, delaminations inside the composite laminate are formed during the DCB testing. For the MO 4 specimens, no secondary delaminations are developed during testing.
- In terms of the SERR at the crack initiation load, the MO 4 is the best performing one. Based on the utilized data reduction scheme, it attains a SERR value of 874 N/m while the worst one MO, the MO 2, achieves a SERR value of 467 N/m.
- "Parasitic" mode mixity is introduced in all DCB tests due to the asymmetry of the joints as well as the presence of residual thermal stresses. Nevertheless, at the load levels that crack initiation occurs, the mode mixity is very low in all four MO examined, lower than 6.1%.

The findings of the present experimental investigation on the fracture behavior of the under-study, novel, titanium-CFRP adhesive joint produced following four different manufacturing techniques would be useful for various high-end applications in the aerospace industry, for instance in wing design using the hybrid laminar flow control technology [17]. Enhancement of the present work with tests using the ENF configuration is presented in our recent publication [21]. Further research will be carried out to investigate the effects of the environmental conditions and fatigue on the fracture behavior of the joint.

# CrediT authorship contribution statement

**Theodoros Loutas:** Conceptualization, Methodology, Writing - review & editing, Supervision, Project administration, Funding acquisition. **Panayiotis Tsokanas:** Methodology, Validation, Formal analysis, Investigation, Data curation, Writing - original draft, Writing - review & editing, Visualization. **Vassilis Kostopoulos:** Conceptualization, Project administration, Funding acquisition. **Peter Nijhuis:** Investigation, Resources. **Wouter M. van den Brink:** Investigation, Resources.

## Declaration of competing interest

The authors declare that they have no known competing financial interests or personal relationships that could have appeared to influence the work reported in this paper.

## Acknowledgments

The work presented in this paper was financially supported by the Clean Sky 2 Joint Undertaking under the European Union's Horizon 2020 research and innovation program TICOAJO (Grant Agreement Number: 737785). The support is appreciated by the authors. Also, the authors thank their TICOAJO partners from the Structural Integrity and Composites (SI&C) research group, Delft University of Technology, The Netherlands, and especially Dr. W. Wang, Prof. J.A. Poulis, Prof. S. Teixeira de Freitas, and Prof. D. Zarouchas, for performing the surface pre-treatment studies. Last, the authors thank their colleagues from the Laboratory of Applied Mechanics & Vibrations, University of Patras, Greece, and especially Mr. D. Pegkos and Dr. G. Sotiriadis, for assisting with the execution of part of the experiments.